\documentclass[runningheads]{svmult}

\usepackage{makeidx}   % allows index generation
\usepackage{graphicx}  % standard LaTeX graphics tool
\usepackage{subeqnar}  % subnumbers individual equations
\usepackage{multicol}  % used for the two-column index
\usepackage{physprbb}  % modified textarea for proceedings,
\makeindex             % used for the subject index
\begin{document}
\title*{On the Way to a Gutzwiller Density Functional Theory}
\toctitle{On the Way to a Gutzwiller Density Functional Theory}
% allows explicit linebreak for the table of content
%
%
\titlerunning{On the Way to a Gutzwiller Density Functional Theory}
% allows abbreviation of title, if the full title is too long
% to fit in the running head
%
\author{Werner Weber\inst{1}
\and J\"{o}rg B\"{u}nemann\inst{2}
\and Florian Gebhard\inst{2}}
\authorrunning{Werner Weber, J\"{o}rg B\"{u}nemann, and Florian Gebhard}
% if there are more than two authors,
% please abbreviate author list for running head
%
%
\institute{Institut~f\"{u}r~Physik, Universit\"{a}t Dortmund, 
D-44221 Dortmund, Germany
\and Fachbereich Physik, Philipps--Universit\"{a}t Marburg, 
D-35032 Marburg, Germany}

\maketitle              % typesets the title of the contribution

%{\bf preprint version of November 3, 2000}
\begin{abstract}
Multi-band Gutzwiller-correlated wave functions   % 6 
reconcile the contrasting concepts of itinerant band electrons % 14
versus electrons localized in partially filled atomic shells. % 22 
The exact evaluation of these variational ground states % 30 
in the limit of large coordination number allows the % 39
identification of quasi-particle band structures, % 45
and the calculation of a variational spinwave dispersion. % 53
The study of a generic two-band model elucidates % 62
the co-operation of the Coulomb repulsion and the % 71
Hund's-rule exchange for itinerant ferromagnetism.  % 77
We present results of calculations for ferromagnetic nickel, % 85
using a realistic 18 spin-orbital basis of $4s$, $4p$ and $3d$ valence % 98 
electrons. The quasi-particle energy bands agree much better % 107
with the photo-emission and Fermi surface data than the band structure % 119
obtained from spin-density functional theory (SDFT). % 125
\end{abstract}

\section{Exchange versus Correlations}

More than 50~years ago two basically different scenarios
had emerged from early quantum-mechanical considerations
on electrons in metals with partly filled $d$~bands.
\begin{description}
\item[Scenario~I:] As proposed by Slater~\cite{Slaterearly} and 
Stoner~\cite{Stoner}, band theory alone was argued to account for itinerant
ferromagnetism. Due to the Pauli principle, 
electrons with parallel spins cannot come
arbitrarily close to each other (``Pauli'' or ``exchange hole''), 
and, thus, a ferromagnetic alignment of the electron spins
reduces the total Coulomb energy with respect to the
paramagnetic situation (``exchange field energy'').
\item[Scenario~II:] As emphasized by van Vleck~\cite{vanVleck},
electronic correlations are important in narrow-band materials.
Due to the strong electron-electron interaction,
charge fluctuations in the atomic $d$~shells are strongly suppressed
(``minimum polarity model'').
The atomic magnetic moments arise due to
the local Coulomb interactions (in particular, 
Hund's-rule couplings) and the electrons' motion through the crystal
may eventually align them at low enough temperatures.
\end{description}
In principle, such a dispute can be resolved in natural sciences.
The corresponding theories have to be worked out in detail, and their
results and predictions have to be compared to experiments.

This was indeed done for scenario~I~\cite{Moruzzi,thisvolume}.
The (spin-)density functional theory
is a refined band theory which describes some
iron group metals with considerable
success. Unfortunately, progress for scenario~II was much slower.
It calls for a theory of correlated electrons, i.e.,
a genuine many-body problem has to be solved.
It was only recently that reliable theoretical tools became available
which allow to elucidate scenario~II in more 
detail~\cite{Nolting,Hasegawa,Pruschke,Vollhardt,BGWvoll,correlatedpeople}.

A first step in this direction was the
formulation of appropriate model Hamiltonians which
allowed to discuss matters concisely, e.g., the Hubbard 
model~\cite{GutzPRL63,Gutzwiller1964,Hubbard,Kanamori}.
This model covers both aspects of $d$~electrons on a lattice:
they can move through the crystal, and they strongly interact
when they sit on the same lattice site.
The model is discussed in more detail in Sec.~\ref{Hubbardmodel}.

Even nowadays, it is impossible to calculate exact ground-state properties
of such a model in three dimensions. 
In 1963/1964 Gutzwiller introduced a 
trial state to examine variationally the possibility of ferromagnetism
in such a model~\cite{GutzPRL63,Gutzwiller1964}. 
His wave function covers both limits of
weak and strong correlations and should, therefore, be suitable
to provide qualitative insights into the magnetic phase diagram
of the Hubbard model. 
Gutzwiller-correlated wave functions for multi-band Hubbard models
are defined and analyzed in Sec.~\ref{GWF}.

The evaluation of multi-band Gutzwiller
wave functions itself poses a most difficult many-body problem.
Perturbative treatments~\cite{Stollhoff,Carmelo} are constrained
to small to moderate interaction strengths. The region of
strong correlations could only be addressed within the
so-called
``Gutzwiller approximation''~\cite{GutzPRL63,Gutzwiller1964,Vollirev}
and its various extensions~\cite{Chao,Fazekas}.
Some ten years ago, the Gutzwiller approximation was found
to become exact for the one-band Gutzwiller wave function
in the limit of infinite spatial dimensions, 
$d\to\infty$~\cite{MVGUTZ,MVPRLdinfty,BUCH}, and
Gebhard~\cite{Geb1990} developed
a compact formalism which allows the straightforward calculation
of the variational
ground-state energy in infinite dimensions.
Recently, Gebhard's approach was generalized by us 
to the case of multi-band
Gutzwiller wave functions~\cite{BGWvoll}. Thereby, earlier results
by B\"unemann and Weber~\cite{BueWeb}, based on 
a generic extension of the Gutzwiller
approximation~\cite{JoergEJB}, were found to 
become exact in infinite dimensions~\cite{BGWhalb}.

As shown in Sect.~\ref{toymodel} for a two-band toy model,
the Gutzwiller variational scheme approach also allows
the calculation of spinwave spectra~\cite{Buene2000}.
In this way, the dispersion relation of the
fundamental low-energy excitations can be derived
consistently. Albeit the description is based on
itinerant electrons, the results for strong ferromagnets
resemble those of a Heisenberg model for localized
spins whereby a unified description of localized and itinerant
aspects of electrons in transition metals is achieved.

In Sect.~\ref{nickel} we discuss results from
a full-scale calculation for nickel. 
The additional local correlations introduced in the Gutzwiller scheme
lead to a much better description of the quasi-particle 
properties of nickel than in previous 
calculations based on spin-density functional theory.

\section{Hamilton Operator}
\label{Hubbardmodel}

Our multi-band Hubbard model~\cite{Hubbard} is defined by the Hamiltonian 
\begin{equation}
\hat{H}=\sum_{i,j;\mathbf{\sigma},\mathbf{\sigma'}}
t_{i,j}^{\mathbf{\sigma},\mathbf{\sigma'}}
\hat{c}_{i;\mathbf{\sigma}}^{+}
\hat{c}_{j;\mathbf{\sigma'}}^{\vphantom{+}}
+ \sum_i\hat{H}_{i;{\rm at}}\equiv \hat{H}_1+\hat{H}_{{\rm at}}\;.  \label{1}
\end{equation}
Here, $\hat{c}_{i;\mathbf{\sigma}}^{+}$ creates an electron with combined
spin-orbit index~$\mathbf{\sigma}=1,\ldots ,2N$ ($N=5$ for 3$d$~electrons) at
the lattice site~$i$ of a solid. 

The most general case is treated in Ref.~\cite{BGWvoll}.
In this work we assume for simplicity that different types
of orbitals belong
to different representations of the point group of the respective 
atomic state (e.g., $s$, $p$, $d(e_g)$, $d(t_{2g})$).
In this case, different types of orbitals do not mix locally, and, thus,
the local crystal field is of the from
$t_{i,i}^{\mathbf{\sigma},\mathbf{\sigma'}}= \epsilon_{\mathbf{\sigma}}
\delta_{\mathbf{\sigma},\mathbf{\sigma'}}$. 
Consequently, we may later work with
normalized single-particle product states $|\Phi_0\rangle$ 
which respect the symmetry of the lattice, i.e.,
\begin{equation}
\langle \Phi_0 | \hat{c}_{i;\mathbf{\sigma}}^{+}
\hat{c}_{i;\mathbf{\sigma'}}^{\vphantom{+}} | \Phi_0 \rangle
= \delta_{\mathbf{\sigma}, \mathbf{\sigma'}} n_{i;\mathbf{\sigma}}^0\; .
\label{eq:2}
\end{equation}
We further assume that the local interaction
is site-independent
\begin{equation}
\hat{H}_{i;{\rm at}} =
\sum_{\mathbf{\sigma_1},\mathbf{\sigma_2},\mathbf{\sigma_3},\mathbf{\sigma_4}}
{\cal U}^{\mathbf{\sigma_1},\mathbf{\sigma_2};\mathbf{\sigma_3},\mathbf{\sigma_4}}
\hat{c}_{i;\mathbf{\sigma_1}}^{+}\hat{c}_{i;\mathbf{\sigma_2}}^{+}
\hat{c}_{i;\mathbf{\sigma_3}}^{\vphantom{+}}
\hat{c}_{i;\mathbf{\sigma_4}}^{\vphantom{+}}\;.  \label{fullHat}
\end{equation}
This term represents all possible local Coulomb interactions.

As our basis for the atomic problem we choose the configuration states
\begin{equation}
|I\rangle = |\mathbf{\sigma_1},\mathbf{\sigma_2}, \ldots\rangle
= \hat{c}_{i;\mathbf{\sigma_1}}^{+}\hat{c}_{i;\mathbf{\sigma_2}}^{+} \cdots
|{\rm vacuum}\rangle \quad (\mathbf{\sigma_1}<\mathbf{\sigma_2}< \cdots)
\; ,
\end{equation}
which are the ``Slater determinants'' in atomic physics.
The diagonalization of the Hamiltonian~$\hat{H}_{i;{\rm at}}$
is a standard exercise~\cite{Sugano}.
The eigenstates~$|\Gamma\rangle$ obey
\begin{equation}
|\Gamma \rangle =\sum_I T_{I,\Gamma }|I\rangle \;,
\end{equation}
where $T_{I,\Gamma }$ are the elements of the unitary matrix which
diagonalizes the atomic Hamiltonian matrix with entries
$\langle I | \hat{H}_{i;{\rm at}}| I'\rangle$.
Then, 
\begin{equation}
\hat{H}_{i;{\rm at}} = \sum_{\Gamma} E_{\Gamma}\hat{m}_{\Gamma} \quad ,
\quad \hat{m}_{\Gamma} = |\Gamma\rangle\langle \Gamma | \; .
\end{equation}
The atomic properties, i.e., eigenenergies $E_{\Gamma}$, 
eigenstates~$|\Gamma\rangle$, and
matrix elements~$T_{I,\Gamma}$, are essential ingredients of our 
solid-state theory.

\section{Multi-band Gutzwiller Wave Functions}
\label{GWF}

\subsection{Variational Ground-State Energy}

Gutzwiller-correlated wave functions 
are written as a many-particle correlator~$\hat{P}_{{\rm G}}$ acting
on a normalized single-particle product state~$|\Phi_0\rangle $, 
\begin{equation}
|\Psi_{{\rm G}}\rangle =\hat{P}_{{\rm G}}|\Phi_0\rangle \;.
\end{equation}
The single-particle wave function $|\Phi_0\rangle$ which obeys~(\ref{eq:2})
contains many configurations which are
energetically unfavorable with respect to the atomic interactions. 
Hence, the correlator~$\hat{P}_{{\rm G}}$ is chosen to 
suppress the weight of these configurations to minimize the total 
ground-state energy of~(\ref{1}). 
In the limit of strong correlations the Gutzwiller 
correlator~$\hat{P}_{{\rm G}}$ 
should project onto atomic eigenstates. Therefore, the proper multi-band
Gutzwiller wave function with atomic correlations reads 
\begin{eqnarray}
\hat{P}_{{\rm G}} &=& \prod_i\hat{P}_{i;{\rm G}} \; , \nonumber \\
\hat{P}_{i;{\rm G}} 
&=&
\prod_{\Gamma} \lambda_{i;\Gamma }^{\hat{m}_{i;\Gamma}}
=\prod_{\Gamma} \left[ 1+\left( \lambda_{i;\Gamma }-1\right) \hat{m}_{i;\Gamma}
\right]  \label{GutzcorrdegbandsHund} 
%\\
%
%&=& 
= 1+\sum_{\Gamma} 
\left( \lambda_{i;\Gamma}-1\right) \hat{m}_{i;\Gamma }
\;. 
%\nonumber 
\end{eqnarray}
The $2^{2N}$ variational parameters~$\lambda_{i;\Gamma }$ per site are
real, positive numbers. For $\lambda_{i;\Gamma_0}\neq 0$ and all other 
$\lambda_{i;\Gamma }= 0$ all atomic configurations at 
site~$i$ but~$|\Gamma_0\rangle $ are removed from~$|\Phi_0\rangle $. 
Therefore, by construction, $|\Psi_{\rm G}\rangle$ covers both limits
of weak and strong coupling. In this way it incorporates both itinerant
and local aspects of correlated electrons in narrow-band systems.

The class of Gutzwiller-correlated wave functions as specified 
in~(\ref{GutzcorrdegbandsHund}) was evaluated exactly in the limit of
infinite dimensions in Ref.~\cite{BGWvoll}. 
The expectation value of the Hamiltonian~(\ref{1}) reads~\cite{hint}
\begin{eqnarray}
\langle \hat{H} \rangle &=& 
\frac{\langle \Psi_{\rm G} | \hat{H} |\Psi_{\rm G} \rangle }%
{\langle \Psi_{\rm G} | \Psi_{\rm G} \rangle } \label{allresultsdegbandsHund}\\
&=&
\sum_{i\neq j;\mathbf{\sigma},\mathbf{\sigma'}}
t_{i,j}^{\,\mathbf{\sigma},\mathbf{\sigma'}} \sqrt{q_{i;\mathbf{\sigma}}}
\sqrt{q_{j;\mathbf{\sigma'}}}
\langle \Phi_0  |
\hat{c}_{i;\mathbf{\sigma}}^{+}
\hat{c}_{j;\mathbf{\sigma'}}^{\vphantom{+}}
| \Phi_0 \rangle 
+\sum_{i;\mathbf{\sigma}}\epsilon_{\mathbf{\sigma}}n_{i;\mathbf{\sigma}}^0
+\sum_{i;\Gamma }E_{\Gamma}m_{i;\Gamma }\;. 
\nonumber 
\end{eqnarray}
Here, $n_{i,\mathbf{\sigma}}^0=\langle \Phi_0 | \hat{n}_{i;\mathbf{\sigma}}|
\Phi_0\rangle $ is the local particle density
in $| \Phi_0 \rangle$. 
The local $q$~factors are given by~\cite{BGWvoll}
\begin{eqnarray}
\sqrt{q_{\mathbf{\sigma}}}  &=& 
\sqrt{\frac{1}{n_{\mathbf{\sigma}}^{0}(1-n_{\mathbf{\sigma}}^{0})}} 
\sum_{\Gamma,\Gamma'} 
\sqrt{\frac{m_{\Gamma}m_{\Gamma'}}{m_{\Gamma}^0m_{\Gamma'}^0}}  
\sum_{ I,I' \, (\mathbf{\sigma}\not\in I,I')} 
f_{\mathbf{\sigma}}^I f_{\mathbf{\sigma}}^{I'} 
 \sqrt{ m_{(I'\cup\mathbf{\sigma})}^{0} m_{I'}^{0} } \nonumber \\
&& 
\hphantom{\sqrt{\frac{1}{n_{\mathbf{\sigma}}^{0}(1-n_{\mathbf{\sigma}}^{0})}} 
\sum_{\Gamma,\Gamma'} 
}
\times T_{\Gamma,(I\cup\mathbf{\sigma})}^+ T_{(I'\cup\mathbf{\sigma}),\Gamma}
T_{\Gamma',I'}^+ T_{I,\Gamma'} \; , \label{qfactor}
\end{eqnarray}
where $m_{i;I}^{0}$ ($m_{i;\Gamma}^{0}$)
is the probability to find the configuration~$|I\rangle$ (the atomic
eigenstate~$|\Gamma\rangle$)
on site~$i$ in the single-particle product state~$|\Phi_0\rangle$.
The fermionic sign function
%\begin{equation}
$f_{\mathbf{\sigma}}^{I}  \equiv \langle I\cup \mathbf{\sigma} 
| \hat{c}_{\mathbf{\sigma}}^+  | I\rangle$
%\end{equation}
gives a minus (plus) sign if it takes an odd (even) number of anticommutations
to shift the operator $\hat{c}_{\mathbf{\sigma}}^+$
to its proper place in the sequence of electron creation operators
in $|I\cup \mathbf{\sigma} \rangle$.

Eqs.~(\ref{allresultsdegbandsHund}) and~(\ref{qfactor}) show that we may
replace the original variational parameters~$\lambda_{i;\Gamma }$ by their
physical counterparts, the atomic occupancies~$m_{i;\Gamma }$.
They are related by the simple equation~\cite{BGWvoll}
\begin{equation}
m_{i;\Gamma } = \lambda_{i;\Gamma }^2 m_{i;\Gamma }^0 \; .
\label{HundgutzrelationsA}
\end{equation}
The probability for an empty site ($|I|=0$) is obtained
from the completeness condition,
\begin{equation}
m_{i;\emptyset }=1-\sum_{\Gamma\,(|\Gamma |\geq 1)}m_{i;\Gamma }\;.  
\label{completenessgamma}
\end{equation}
The probabilities for a singly occupied site ($|I|=1$) are given by~\cite{hint}
\begin{subeqnarray}
m_{i;\mathbf{\sigma}} &=& n_{i;\mathbf{\sigma}}^0 -
\sum_{I \, (|I|\geq 2) \, (\mathbf{\sigma}\in I) } m_{i;I} \;, 
\label{mImGammastrichA}\\
m_{i;I} &=& \sum_{K} \biggl|
\sum_{\Gamma} 
\sqrt{\frac{m_{i;\Gamma}}{m_{i;\Gamma}^0}}
T_{\Gamma,I}^+ T_{K,\Gamma}
\biggr|^2 m_{i;K}^0 \; .  \label{mImGammastrichB}
\end{subeqnarray}%
The parameters $m_{i;\emptyset}$ and $m_{i;\mathbf{\sigma}}$
must not be varied independently.
All quantities in~(\ref{allresultsdegbandsHund}) are now expressed in
terms of the atomic multi-particle
occupancies~$m_{i;\Gamma}$ ($|\Gamma|\geq 2$),
the local densities $n_{i;\mathbf{\sigma}}^0$, and further variational
parameters in~$|\Phi_0\rangle$.

It is seen that the variational ground-state energy can be cast into the
form of the expectation value of an effective single-particle Hamiltonian
with renormalized electron transfer amplitudes
$\widetilde{t}_{i,j}^{\,\mathbf{\sigma},\mathbf{\sigma'}}$,
\begin{eqnarray}
\hat{H}_{\rm eff} &=& 
\sum_{i\neq j;\mathbf{\sigma},\mathbf{\sigma'}}
\widetilde{t}_{i,j}^{\,\mathbf{\sigma},\mathbf{\sigma'}} 
\hat{c}_{i;\mathbf{\sigma}}^{+}
\hat{c}_{j;\mathbf{\sigma'}}^{\vphantom{+}}
+\sum_{i;\mathbf{\sigma}}\epsilon_{\mathbf{\sigma}}\hat{n}_{i;\mathbf{\sigma}}
+\sum_{i;\Gamma }E_{\Gamma}m_{i;\Gamma }\, , \nonumber \\
\widetilde{t}_{i,j}^{\,\mathbf{\sigma},\mathbf{\sigma'}} 
&=& 
\sqrt{q_{i;\mathbf{\sigma}}}
\sqrt{q_{j;\mathbf{\sigma'}}}
t_{i,j}^{\,\mathbf{\sigma},\mathbf{\sigma'}}  \; .
\label{widetildet}
\label{Heff}
\end{eqnarray}
Therefore, $|\Phi_0\rangle$ is the ground state of $\hat{H}_{\rm eff}$
whose parameters have to be determined self-consistently
from the minimization of $\langle \Phi_0| \hat{H}_{\rm eff}| \Phi_0 \rangle$ 
with respect to $m_{i;\Gamma}$ and $n_{i;\mathbf{\sigma}}^0$.
For the optimum set of parameters, $\hat{H}_{\rm eff}^{\rm opt}$ 
{\sl defines\/} a band structure for {\sl correlated\/} electrons. 
Similar to density-functional theory, this interpretation
of our ground-state results opens the way to detailed comparisons with 
experimental results; see Sect.~\ref{CompEXp}.

\subsection{Spinwaves}
\label{spinwave}

The variational principle can also be used to calculate
excited states~\cite{Messiah}. 
If $|\Phi\rangle$ is the ferromagnetic, exact
ground state with energy $E_0$, the trial states
\begin{equation}
|\Psi(q) \rangle = \hat{S}_q^- |\Phi\rangle
\end{equation}
are necessarily orthogonal to $|\Phi\rangle$, and provide
an exact upper bound to the first excited state with momentum~$q$
and energy $\epsilon(q)$
\begin{equation}
\epsilon(q) 
\leq E_{\rm s}(q) \equiv 
\frac{\langle \Psi(q) | \hat{H} | \Psi(q) \rangle}%
{\langle \Psi(q) | \Psi(q) \rangle} - E_0 \;.
\end{equation}
Here, $\hat{S}_q^-= (\hat{S}_q^+)^+= \sum_{l,b} \exp(-\I ql) 
\hat{c}_{l,b,\downarrow}^+\hat{c}_{l,b,\uparrow}^{\vphantom{+}}$
flips a spin from up to down in the system whereby it changes
the total momentum of the system by~$q$.
In this way, the famous Bijl-Feynman formula for 
the phonon-roton dispersion in superfluid Helium
was derived~\cite{Feynman}.
In the case of ferromagnetism the excitation energies~$E_{\rm s}(q)$ 
can be identified with
the spinwave dispersion if a well-defined spinwave exists at 
all~\cite{Buene2000}.
Experimentally this criterion is fulfilled for small momenta~$q$ 
and energies~$E_{\rm s}(q)$.

Unfortunately, we do not know the exact ground state or its energy
in general.
However, we may hope that the Gutzwiller wave function~$|\Psi_{\rm G}\rangle$
is a good approximation to the true ground state.
Then, the states
\begin{equation}
|\Psi_{\rm G}(q) \rangle = \hat{S}_q^- |\Psi_{\rm G}\rangle
\end{equation}
will provide a reliable estimate for $E_{\rm s}(q)$,
\begin{equation}
E_{\rm s}(q) \approx E_{\rm s}^{\rm var} (q)= 
\frac{
\langle \Psi_{\rm G} | \hat{S}_q^+ \hat{H} \hat{S}_q^-| \Psi_{\rm G} \rangle
}{
\langle \Psi_{\rm G} | \hat{S}_q^+ \hat{S}_q^-| \Psi_{\rm G} \rangle
}
-
\frac{
\langle \Psi_{\rm G} | \hat{H} | \Psi_{\rm G} \rangle
}{
\langle \Psi_{\rm G} | \Psi_{\rm G} \rangle
}
\;.
\label{eq:spinwavedispersion}
\end{equation}
Naturally, $E_{\rm s}^{\rm var} (q)$ does not obey any strict 
upper-bound principles.

The actual calculation of the variational spinwave dispersion is
rather involved. However, explicit formulae are available~\cite{Buene2000}
which can directly be applied once the variational parameters have
been determined from the minimization of the variational ground-state energy.

\begin{figure}[htb]
\vspace*{-15mm}
  \begin{center}
    \includegraphics[width=8cm,height=12cm]{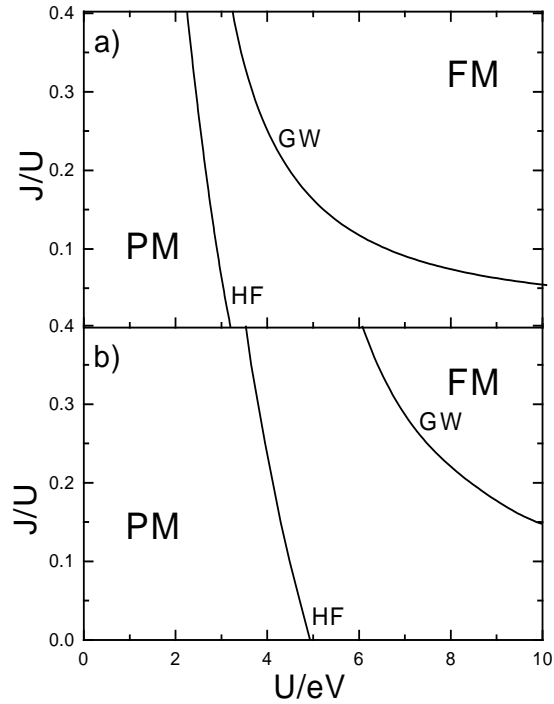}
  \end{center}
\vspace*{-13mm}
\caption{Phase diagram as a function of~$U$ and~$J$ for the
Hartree--Fock--Stoner theory (HF) and the Gutzwiller wave function
(GW) for \textbf{(a)}~$n=1.17$ and 
\textbf{(b)}~$n=1.40$; 
PM: paramagnet, FM: ferromagnet}
\label{fig:phasediaferro}
\end{figure}

\section{Results for a Generic Two-Band Model}
\label{toymodel}

\subsection{Ground-State Properties}

The atomic Hamiltonian for a two-band model ($b=1,2$)
can be cast into the form 
\begin{eqnarray}
\hat{H}_{i;{\rm at}} &=&
U \sum_{b}\hat{n}_{b,\uparrow}\hat{n}_{b,\downarrow}
+U'\sum_{\sigma,\sigma'}\hat{n}_{1,\sigma}\hat{n}_{2,\sigma'}
-J\sum_{\sigma}\hat{n}_{1,\sigma}\hat{n}_{2,\sigma}
\nonumber 
\\[3pt]
&& +J\sum_{\sigma}\hat{c}_{1,\sigma}^{+}
\hat{c}_{2,-\sigma}^{+}
\hat{c}_{1,-\sigma}^{\vphantom{+}}
\hat{c}_{2,\sigma}^{\vphantom{+}}  
+J_{{\rm C}} \Bigl(
\hat{c}_{1,\uparrow}^{+}\hat{c}_{1,\downarrow}^{+}
\hat{c}_{2,\downarrow}^{\vphantom{+}}\hat{c}_{2,\uparrow}^{\vphantom{+}}
+ 
\hat{c}_{2,\uparrow}^{+}\hat{c}_{2,\downarrow}^{+}
\hat{c}_{1,\downarrow}^{\vphantom{+}}\hat{c}_{1,\uparrow}^{\vphantom{+}}
\Bigr)\;. \label{twoorbhamiltonian}
\end{eqnarray}
For two $d(e_g)$~orbitals, $\hat{H}_{{\rm at}}$ exhausts all possible
two-body interaction terms.
Since we assume that the model describes two degenerate $d(e_g)$ orbitals,
the following restrictions are enforced by the cubic 
symmetry~\cite{Sugano}:
(i)~$J=J_{{\rm C}}$, and (ii)~$U-U'=2J$.
Therefore, there are two independent Coulomb parameters,
the local Coulomb repulsion $U$ (of the order of $10\, {\rm eV}$)
and the local exchange coupling~$J$ (of the order of $1\, {\rm eV}$,
as typical for atomic Hund's rule couplings).
For the one-particle part~$\hat{H}_1$ we 
use an orthogonal tight-binding Hamiltonian with first and second
nearest neighbor hopping matrix elements, 
resulting in a bandwidth~$W=6.6\, {\rm eV}$.

In the following we concentrate on two band-fillings, (a),
$n=1.17$, where the non-interacting density of states (DOS) shows a 
pronounced peak at the Fermi energy, 
most favorably for ferromagnetism, and, (b),
$n=1.40$, a position near the DOS peak, where the DOS exhibits a positive
curvature as a function of the magnetization.

In Fig.~\ref{fig:phasediaferro} 
we display the $J$-$U$ phase diagram for both
fillings. It shows that Hartree--Fock theory always predicts a ferromagnetic
instability. In contrast, the correlated-electron approach strongly supports
the ideas of van Vleck~\cite{vanVleck} and Gutzwiller~\cite{Gutzwiller1964}:
(i)~a substantial on-site exchange~$J$ is required for the occurrence
of ferromagnetism if, (ii), realistic Coulomb repulsions~$U$
are assumed. 
At the same time the comparison of Figs.~\ref{fig:phasediaferro}a
and~\ref{fig:phasediaferro}b shows the importance of
band-structure effects which are the basis of the Stoner theory.
The ferromagnetic phase in the $U$-$J$ phase diagram is much bigger when
the density of states at the Fermi energy is large. Therefore, the
Stoner mechanism for ferromagnetism is well taken into account
in our correlated-electron approach.

\begin{figure}[htb]
\vspace*{-16mm}
  \begin{center}
    \includegraphics[width=8cm]{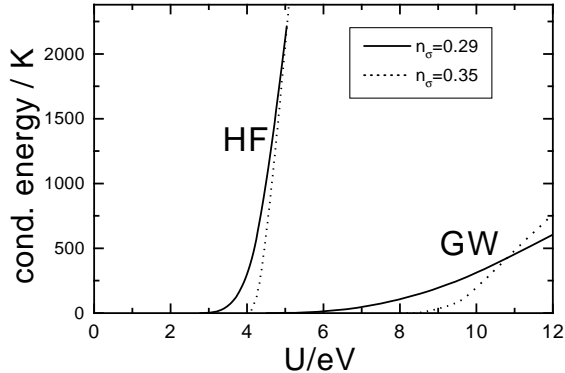}
  \end{center}
\vspace*{-52mm}
\caption{Condensation energy as a function of~$U$ for $J=0.2U$ for the
Hartree--Fock theory (HF) and the Gutzwiller wave function 
(GW) for $n=1.17$ (full lines) 
and~$n=1.40$ (dashed lines)}
\label{fig:condenenergy}
\end{figure}

In Fig.~\ref{fig:condenenergy}, we display the energy differences
between the paramagnetic and ferromagnetic ground states 
(``condensation energy'', $E_{\rm cond}$) as a function of
the interaction strength for $J=0.2U$. 
This quantity should be of the order of the Curie temperature which is in the
range of $100\,{\rm K}$--$1000\,{\rm K}$ in real materials. 
The Hartree--Fock--Stoner theory yields 
such small condensation energies only in the range 
of $U\approx 4\,$eV; for larger $U$, $E_{\rm cond}$ 
is of order $U$. In any case, the interaction parameter~$U$ 
has to be tuned very precisely to give condensation energies
which concur with experimental Curie temperatures~\cite{Slaterearly}.
In contrast, for the Gutzwiller-correlated wave function, we find
relatively small condensation energies $E_{\rm cond}=0.5\cdot 10^3\, {\rm K}$
even for interaction values
as large as twice the bandwidth ($U\approx 12\,$eV).
Moreover, the dependence of the condensation energy on~$U$ is rather weak
such that uncertainties in~$U$ do not drastically influence the
estimates for the Curie temperature.

\subsection{Spinwave Dispersions}
\label{spinwavedisP}

\begin{figure}[ht]
\vspace{-5mm}
  \begin{center}
   \includegraphics[width=7.1cm]{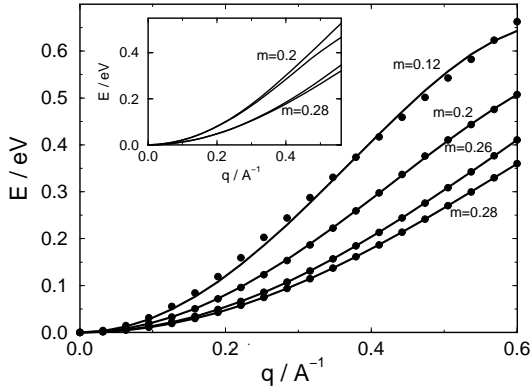}
  \end{center}
\vspace{-4mm}
\caption{Variational spinwave dispersion in $x$~direction,
$E_{\rm s}^{\rm var}((q,0,0))$,
for the two-band model defined in Sect.~\ref{toymodel};
$n=1.17$, $J=0.2U$, and the values 
$U/{\rm eV}=7.8,10,12,13.6$ correspond to $m=0.12,0.20,0.26,0.28$.
The lattice constant is $a=2.5\,$\AA. Inset: 
$E_{\rm s}^{\rm var}((q,0,0))$ and 
$E_{\rm s}^{\rm var}((q/\sqrt{2},q/\sqrt{2},0))$
for $m=0.2$ and $m=0.28$, respectively. The spinwave dispersion is 
almost isotropic}
\label{fig:spinwaves}
\end{figure}

In Fig.~\ref{fig:spinwaves} we show $E_{\rm s}^{\rm var}((q,0,0))$,
the variational spinwave dispersion~(\ref{eq:spinwavedispersion}),
in $x$~direction
for the model parameters $n=1.17$, 
$J=0.2U$, and the four different
values $U/{\rm eV}=7.8,10,12,13.6$ which
correspond to a magnetization per band 
of $m=0.12,0.20,0.26,0.28$.
This quantity is defined as 
$0 \leq m=(n_{b,\uparrow}-n_{b,\downarrow})/2\leq n/4$.
Note that our last case corresponds to an almost complete 
ferromagnetic polarization.
The data fit very well the formula
\begin{equation}
E_{\rm s}^{\rm var}((q,0,0)) = Dq^2(1+\beta q^2) +{\cal O}(q^6) \; ,
\end{equation}
in qualitative agreement with experiments on nickel~\cite{nickelexp}.
The corresponding values $D=1.4\, {\rm eV}\,$\AA$^2$ and 
$D=1.2\, {\rm eV}\,$\AA$^2$ 
for $m=0.26$ and $m=0.28$, respectively, 
are of the right order of magnitude for nickel
where $D=0.43\, {\rm eV}\,$\AA$^2$. As lattice constant of
our simple-cubic lattice we chose $a=2.5\,$\AA.

As shown in the inset of Fig.~\ref{fig:spinwaves},
the dispersion relation is almost isotropic for
$q$~values up to half the Brillouin zone boundary~\cite{Buene2000},
in particular for large magnetizations. This is in contrast to
the strong dependence of the electron-transfer amplitudes~$t_{i,j}$
on the lattice direction. This implies for {\sl strong\/} ferromagnets
that the collective motion
of the local moments is similar to that of {\sl localized\/} spins
in an {\sl insulator}~\cite{Eschrig}. 
Such ferromagnetic insulators are conveniently
described by the Heisenberg model with exchange interactions between
neighboring sites $\langle i,j\rangle$ on a cubic lattice,
\begin{equation}
\hat{H}_{\rm S} = - J \sum_{\langle i,j\rangle} \hat{\vec{S}}_i\hat{\vec{S}}_j
\; .
\end{equation}
For such a model one finds $D=2S J a^2$.
The length of the effective
local spins can be calculated from $|\Psi_{\rm G}\rangle$ 
as $S(S+1)\approx 0.95$
($S=0.6$) for $m\geq 0.20$~\cite{BGWvoll}. 
Therefore, $J\approx D/(1.2 a^2) $, which gives
the typical value $J= 0.17\, {\rm eV}$. For an estimate of the Curie
temperature~$T_{\rm C}$ we use the result from quantum Monte-Carlo
calculations~\cite{QMCspinmodels}
\begin{equation}
T_{\rm C} = 1.44 J S^2 
\end{equation}
for spins~$S$ on a simple-cubic lattice.
In this way we find $T_{\rm C}\approx 0.5 J=0.09\, {\rm eV}=
1\cdot 10^3 \, {\rm K}$.
This is the same order of magnitude as the condensation energy for these
values of the interaction, $E_{\rm cond} = 5 \cdot 10^2 \, K$,
see Sect.~\ref{toymodel}.
Given the arbitrariness in the relation between
$E_{\rm cond}$ and $T_{\rm C}$, and the application of the Heisenberg model
to our itinerant-electron system, we may certainly allow for 
difference of a factor two in these quantities.
Nevertheless, the results of this section clearly show that, (i),
$E_{\rm cond}$ gives the right order of magnitude for $T_{\rm C}$, and that,
(ii), the spinwave dispersion of {\sl strong itinerant\/} ferromagnets
resembles the physics of {\sl localized\/} spins.

\section{Correlated Band-Structure of Nickel}
\label{nickel}

\subsection{Discrepancies Between Experiment and SDFT}

Of all the iron group magnetic metals, nickel is the most celebrated case
of discrepancies between the results from experiment and from 
spin-density functional theory (SDFT)~\cite{Huefner}. From very
early on, the photo-emission data have indicated that the width of
the occupied part of the $d$~bands is 
approximately $W^*_{\rm occ} = 3.3\, {\rm eV}$~\cite{Eberhardt} 
whereas all SDFT 
results yield values of $W^*_{{\rm occ, SDFT}}=4.5\, {\rm eV}$ or 
larger~\cite{Moruzzi,Eberhardt}.
Similarly, the low temperature specific heat data~\cite{Dixon} give
a much larger value of $N^*(E_{\rm F})$, 
the quasi-particle density of states at
the Fermi energy ($3.0$ vs.~$1.9$ states/(eV atom)), which indicates a
quasi-particle mass enhancement by a factor of approximately $1.6$.
Here, the Sommerfeld formula is used to convert the specific heat data;
the theoretical value follows directly from the quasi-particle 
band structure. Furthermore, very detailed 
photo-emission studies at symmetry points and
along symmetry lines of the Brillouin zone show discrepancies to SDFT
results for individual band-state energies which are of similar magnitude
as seen in the overall $d$~bandwidth. 

The studies revealed even bigger
discrepancies in the exchange splittings of majority spin and minority
spin bands. The SDFT results give a rather isotropic exchange splitting of
about $600\, {\rm meV}$~\cite{Moruzzi,Eberhardt,Callaway}. In contrast, 
the photo-emission data show
small and highly anisotropic exchange splittings between $160\, {\rm meV}$ 
for pure $d(e_g)$~states such as $X_2$ and $330\, {\rm meV}$ for 
pure $d(t_{2g})$~states, the latter
value estimated from the exchange splitting of $\Lambda_3$ states along
$\Gamma$ to $L$~\cite{Donath,Guenthe}.
The much larger and much too isotropic exchange splitting of the SDFT results
has further consequences.
\begin{enumerate}
\item
The experimental magnetic moment of the strong ferromagnet Ni is
$\mu= 0.61 \mu_{\rm B}$; yet of relevance is its spin-only part
$\mu_{\hbox{\scriptsize spin-only}} = 0.55 \mu_{\rm B}$~\cite{Mook}. 
The SDFT result is $\mu_{\hbox{\scriptsize spin-only}} = 0.59 \mu_{\rm B}$~\cite{Moruzzi},
an overestimate related to the too large exchange splitting.
\item
the $X_2$~state of the minority spin bands lies below 
$E_{\rm F}$~\cite{Hopster}, whereas
all SDFT results predict it to lie above the Fermi 
level~\cite{Moruzzi,WangCallaway,Wohlfahrtreview}. As a consequence,
the SDFT Fermi surface exhibits {\sl two\/} hole ellipsoids 
around the $X$ point of the
Brillouin zone whereas in the de-Haas--van-Alphen experiments only {\sl one\/}
ellipsoid has been found~\cite{WangCallaway,Tsui}.
\item The strong $t_{2g}$-$e_g$ anisotropy is also reflected 
in the total $d$~hole spin density, i.e., in the observation 
that the $d$-hole part of the Ni magnetic
moment has 81\% $d(t_{2g})$ and 19\% $d(e_g)$ character~\cite{Mook}, 
whereas the SDFT results give a ratio of 74\% to 26\%~\cite{Jepsen}.
\end{enumerate}
In the late 70's and early 80's various authors have investigated in how
far many-body effects improve the agreement between theory and 
experiment, see, e.g., Refs.~\cite{Cooke,Liebsch}.
For example, Cooke et al.~\cite{Cooke} introduced an anisotropic exchange
splitting as a fit parameter.

\subsection{Present Status of the Gutzwiller-DFT}
\label{critics}

\subsubsection*{Limitations:}

By construction, the Gutzwiller approach naturally combines
with density-functional theory (DFT) which provides a
basis of one-particle wave functions and a `bare' band structure.
The Gutzwiller-DFT introduces important local correlations
and provides a variational ground-state energy, a
quasi-particle band structure, and a spin-wave dispersion.

Nevertheless, the Gutzwiller-DFT has its own limitations
which we collect here for further reference.
\begin{enumerate}
\item It starts from a model Hamiltonian whose parameters need to be
determined from a DFT calculation; we shall
comment on this procedure below.
\item The true ground state is approximated by a variational many-body
wave function; however, our experience from the two-band model supports
our hope that the variational freedom of our wave function is big enough
to capture the essential features of itinerant ferromagnetism
in real materials as well.
\item The variational ground-state energy is evaluated exactly only
in the limit of infinite dimensions; however, from the one-band case, we expect
$1/d$~corrections to be small~\cite{Geb1990}.
\item Similar in spirit to density-functional theory,
we {\sl interpret\/} the ground-state energy in terms of 
a quasi-particle band structure;
it should be kept in mind, though, that this quantity is, in general,
not identical to the quasi-particle dispersion in the sense of
standard many-body theory~\cite{FetterWalecka}.
\item Most dynamic quantities, e.g., the spectral function,
cannot be determined within our approach;
the example of the spinwave dispersion in Sect.~\ref{spinwavedisP}
shows, however, that we can calculate low-order moments of spectral functions
consistently.
\end{enumerate}
Despite all these restrictions, the method remedies
many problems of the SDFT in the description of
the quasi-particle band structure of nickel, see Sect.~\ref{CompEXp}.

\subsubsection*{Parameterization of the One-Particle Hamiltonian:}

In the present study, we determine the
hopping matrix elements~$t_{i,j}^{\mathbf{\sigma},\mathbf{\sigma'}}$
in~(\ref{1}) from a least squares' fit to the energy bands obtained from
a density-functional-theory calculation for non-magnetic nickel.
An orthogonal nine orbital basis is used, 
and the root-mean-square deviation of the $d$~band energies is about
$60\, {\rm meV}$. 

A more complete description should include the flexibility of the wave 
functions to relax in the magnetic state. This could be achieved by 
enhancing the orbital basis by 4$d$ states. Moreover, 
spin-orbit coupling is of significance in nickel, 
as it leads to a 10\% enhancement of the total magnetic moment. 
In principle, the spin-orbit coupling, or, more generally, an arbitrarily 
large orbital basis can be treated within our formalism~\cite{BGWvoll}, 
yet it leads to complications such as local~$q$~factors which now 
depend on two spin-orbital indices instead of one as in~(\ref{qfactor}). 
These extensions not only enhance the numerical complexity of the problem
but also require different methods for extracting the single-particle 
Hamiltonian from DFT, for example by a more direct evaluation of DFT 
results obtained from local basis methods. 

Since we start from a DFT basis,
the `bare' band structure incorporates already some important
exchange and correlation effects. In particular, we may expect that
the non-local Coulomb terms are well taken into account because
the electron-electron interaction is screened at a length scale of the
order of the inverse Fermi wave number. 
In this way, we can restrict all explicit Coulomb interaction terms
in~$\hat{H}$ to local interactions. 
This assumption is supported by the fact
that the Hartree--Fock approximation becomes exact in infinite
dimensions for density-density interactions,
$\hat{V}^{\mathbf{\sigma},\mathbf{\sigma'}} ({r\neq 0})=
\sum_l \hat{n}_{l,\mathbf{\sigma}}\hat{n}_{l+r,\mathbf{\sigma'}}
\to \hat{V}_{\rm HF}^{\mathbf{\sigma},\mathbf{\sigma'}} 
({r\neq 0})$~\cite{MHart}.
Therefore, we expect that interaction terms beyond the purely
local Hubbard interaction
should be properly taken into account in the density-functional approach
in three dimensions. However, the proper treatment of the ``double counting'' 
problem for both local and non-local interactions 
remains a serious problem for all methods which try to combine 
density-functional approaches with model-based many-particle
theories; see, e.g., the contributions by Lichtenstein, Vollhardt, and
Potthoff in this volume.

\subsubsection*{Chemical Potentials:}

In the translationally invariant system under investigation,
the local occupation densities are the same as their system averages,
\begin{equation}
\langle \hat{n}_{i,\mathbf{\sigma}}\rangle = 
\langle \hat{N}_{\mathbf{\sigma}}\rangle/L\; ,
\end{equation}
where $\hat{N}_{\mathbf{\sigma}}=\sum_i \hat{c}_{i;\mathbf{\sigma}}^{+}
\hat{c}_{i;\mathbf{\sigma}}^{\vphantom{+}}$ counts the number of electrons
with spin-orbit index~$\mathbf{\sigma}$.
Therefore, we may equally work with 
chemical potentials $\mu_{\mathbf{\sigma}}$ for each spin-orbit index
in the Hamiltonian
\begin{equation}
\hat{H}_{\rm gc} = \hat{H} -\sum_{\mathbf{\sigma}} \mu_{\mathbf{\sigma}}
 \hat{N}_{\mathbf{\sigma}}\; .
\end{equation}
In this grand-canonical view, the chemical potentials rather than the
particle densities act as variational parameters.
Naturally, not all of these parameters may be varied independently.
For example, as a consequence of the hybridization of the 4$sp$ and the 3$d$
electrons, the 3$d$ levels would be depleted for a strong
$d$-$d$ repulsion which needs to
be compensated using one of the parameters.
Presently we keep fixed the values of the 4$s$ and 4$p$ partial charges, 
and thus also the 3$d$ total charge, to the values of the non-magnetic 
calculation. This is achieved by using two of the four 
chemical potentials for 4$s$ and 4$p$ electrons.

As can be seen from~(\ref{Heff}), the chemical potentials act as
a shift of the `bare' (DFT) values of the fields $\epsilon_{\mathbf{\sigma}}$,
\begin{equation}
\epsilon_{\mathbf{\sigma}}^{\rm eff} = 
\epsilon_{\mathbf{\sigma}} - \mu_{\mathbf{\sigma}} \; .
\label{chempot}
\end{equation}
In this way, the variational approach naturally contains the flexibility
to adjust the magnetic (or ``exchange'') splitting 
between majority bands $(b,\uparrow)$ and minority bands $(b,\downarrow)$,
\begin{equation}
\Delta_b= \epsilon_{b,{\uparrow}}^{\rm eff}
 -\epsilon_{b,{\downarrow}}^{\rm eff}\; .
\label{Deltadefs}
\end{equation}
In particular, we may allow for an anisotropy in the exchange splittings
of the $d(e_g)$ and $d(t_{2g})$ electrons.

\subsubsection*{Interaction Parameters of the Atomic Hamiltonian:}

Presently we employ only the on-site Coulomb interaction 
within the 3$d$ shell, i.e., all interactions within 
the 4$s$, 4$p$ shell and between 4$sp$ and 3$d$ are neglected. 
In spherical atom approximation, which is found to be well justified, 
all matrix elements in~(\ref{fullHat}) can either be expressed 
as a function of the Slater integrals~$F(k)$ ($k=0,2,4$) or 
of the Racah parameters 
$A$, $B$, $C$~\cite{Sugano}. We use $C/B \approx 4$--$5$~\cite{Sugano}
and determine $A$ and $C$ 
in order to give an optimal agreement with experimental data
(effective mass and bandwidth, condensation energy, 
$t_{2g}/e_g$~ratio of the $d$~part of the magnetic moment,
Fermi surface topology). 

Currently, there is a big debate on the 
magnitude of the interaction parameters.
In principle, the interaction parameters 
could also be deduced from DFT results. 
However, there is no consensus on
how to calculate these parameters consistently.
For example, 
they could be calculated from atomic or Wannier functions,
or they could be found using constrained DFT methods (see, e.g., 
Ref.~\cite{Vielsack}). 

\subsubsection*{Minimization:}

The number of multi-electron
states~$|\Gamma\rangle$ is $2^{2N}=2^{10}$. Because of the cubic site
symmetry, the number of independent variational 
parameters~$m_{\Gamma}$ reduces to approximately~200 for the paramagnetic
and to approximately~400 for the ferromagnetic cases.
These ``internal'' variational parameters obey $2N + 1$ sum 
rules~(\ref{completenessgamma}) and~(\ref{mImGammastrichA});
in cubic symmetry there remain three for the paramagnetic and five for the 
ferromagnetic cases. There is freedom to choose those $m_{\Gamma'}$
which, through the sum rules, are dependent on the other $m_{\Gamma}$. 
It is advisable to pick those $m_{\Gamma'}$ which can be expected 
to have large values. This avoids unphysical negative values of 
$m_{\Gamma'}$ to occur during the variational procedure.

The chemical potentials of~(\ref{chempot}) are the ``external'' 
variational parameters. In the present case these are eight, however 
three are fixed to yield the total 4$s$, 4$p$, and 3$d$ densities, 
such that the space of the external parameters is five-dimensional.
Given a fixed set of external variational parameters, 
the procedure to determine the internal ones begins to put them 
equal to their uncorrelated values $m_{\Gamma}=m^0_{\Gamma}$. 
Thus, $q^0_{d,\sigma} = 1$. 
Note that $q_{s,\sigma} = q_{p,\sigma} = 1$ always holds, 
as there is no interaction for 4$s$, 4$p$ orbitals. 
From this, the `bare' (DFT) band structure and 
$|\Phi_0\rangle^{\rm bare}$ follow as an initial guess
for the quasi-particle band structure and one-particle product state. 
Then, the following self-consistent scheme is employed:
\begin{enumerate}
\item Calculate the ground-state energy for 
$|\Phi_0\rangle_{\alpha}^{\rm old}$ where $\alpha$ labels the set of external 
variational parameters.
This requires momentum-space integrations up to the respective Fermi surface.
\item Minimize the ground-state energy~(\ref{allresultsdegbandsHund})  
with respect to the internal variational parameters.
\item Calculate the $q$~factors and derive 
$|\Phi_0\rangle_{\alpha}^{\rm new}$ as the ground state of
the $\hat{H}_{\rm eff}$~(\ref{Heff}) with the renormalized
hopping matrix elements $\widetilde{t}_{i,j}$;
repeat steps 1--3 until convergence 
to $|\Phi_0\rangle_{\alpha}^{\rm opt}$ is reached.
\end{enumerate}
Self-consistency is usually reached rather quickly, i.e.,
$|\Phi_0\rangle_{\alpha}^{\rm opt}$ is found after three to five iterations.

The global minimum, $|\Phi_0\rangle_{\rm global}^{\rm opt}$
is found by a search through the space of 
the external variational parameters keeping the
average $d$ and $sp$ occupations.
This search can be sped up by first optimizing
with respect to the most important external variational
parameter which is the isotropic exchange splitting~$\Delta_d=
( \Delta_{e_g} + \Delta_{t_{2g}})/2$, putting the difference to zero
as a first approximation.

%{\bf Hi Werner: did we correctly guess what you are doing???}

In a second step, the anisotropy of the exchange splitting
is investigated, i.e., we introduce $\Delta_{e_g}$ and 
$\Delta_{t_{2g}}$ in the minimization procedure, 
keeping $\Delta_d$ at the value
of $\Delta_d^{\rm opt}$ obtained in the first optimization step. 
The searches for $\Delta_d^{\rm opt}$,
and for $\Delta_{e_g}^{\rm opt}$ and 
$\Delta_{t_{2g}}^{\rm opt}$ can be carried out starting
with $|\Phi_0\rangle^{\rm bare}$. Only then the self-consistency
procedure for $|\Phi_0\rangle^{\rm opt}$ has to be launched.

Typical energy gains are (in meV):
\begin{subeqnarray}
E_0^{\rm bare} - E_0^{\rm bare}(\Delta_d^{\rm opt}) 
& \approx & \hbox{$10$--$100$} ,\\
E_0^{\rm bare}(\Delta_d^{\rm opt}) - 
E_0^{\rm bare}(\Delta_{e_g}^{\rm opt},\Delta_{t_{2g}}^{\rm opt})
& \approx & \hbox{$5$--$10$} ,\\
E_0^{\rm bare}(\Delta_{e_g}^{\rm opt},\Delta_{t_{2g}}^{\rm opt})
- E_0^{\rm opt}(\Delta_{e_g}^{\rm opt},\Delta_{t_{2g}}^{\rm opt})
& \approx & \hbox{$5$--$10$}\; .
\end{subeqnarray}%
The energy gains from the variations of 
$\Delta_{s}$ and $\Delta_{p}$ are of the order of $0.1\, {\rm meV}$.

\subsection{Comparison to Experiments}
\label{resultsforNi}
\label{CompEXp}

The results for nickel of our DFT-based Gutzwiller calculations agree
best with experiment when we choose the following values of the
interaction parameters: $A \approx 10$--$12\, {\rm eV}$,
$C \approx 0.1$--$0.4\, {\rm eV}$ with $C/B \approx 4.5$~\cite{Jubelpaper}.
The width of the $d$~bands is predominantly determined by $A$ (essentially
the Hubbard~$U$) via the values of the hopping reduction factors 
$q_{d,\sigma}$.
The exchange splittings and, consequently, the magnetic moment are mainly
governed by~$C$ and to some extend also by~$A$. 
The Racah parameter~$C$ causes the Hund's-rule
splitting of the $d^8$ multiplets; in the hole picture, $d^8$ is the only 
many-particle configuration which is significantly 
occupied (by $1.90$~electrons),
while $5.94$~electrons are in $d^9$, $0.89$~electrons are in $d^{10}$,
and $1.18$~electrons have $s$~or $p$~character.

In our present study, the parameter~$C$ is found 
to be rather small ($0.1\, {\rm eV}$) 
compared to~$A$ in order to reproduce the measured 
spin-only moment~$\mu_{\hbox{\scriptsize spin-only}} = 0.55$.
Larger values of~$C$ move the minimum of the total
energy curve $E_{\rm tot}$ vs.~magnetization~$m$ to values of
$m\approx 0.60$--$0.65 \mu_{\rm B}$.

There are two points to discuss here.
The first concerns the large value of~$A$, which seems incompatible with
the position of the satellite peak in the photo-emission data at
about $6\, {\rm eV}$ below the Fermi energy~$E_{\rm F}$~\cite{Huefner}.
Model calculations for this many-body excitation
peak use values of $U \approx 3$--$5\, {\rm eV}$. 
However, these models use
single of few $d$~band models, excluding hybridization with the 4$s$, 4$p$ 
bands, see, e.g., Ref.~\cite{Liebsch}.
When, in our calculation, the hybrization effects
are switched off, and only the $d$~band contribution to the total energy
matters, we also find that values of $A \approx 3$--$5\, {\rm eV} $
agree best with experiment, and $A \approx 10\, {\rm eV}$ 
would be way out of a reasonable range of parameter values.

The second point concerns the shape of the total energy curve 
$E_{\rm tot}(m)$ at large values of~$m$ in the limit of strong
ferromagnetism. In this limit, the increase of the magnetic moment
is fed from the $d$~admixture in the majority 4$s$, 4$p$ bands. Compared
to analogous curves obtained from SDFT, the curvature at large~$m$ values
is much smaller in our results. We presume that the larger SDFT curvature
is related to the balance between 4$s$, 4$p$ and 3$d$ electrons. 
It is well known
that this balance in a delicate manner determines the stability of
transition metals as well as of noble metals;
see, e.g., Ref.~\cite{Pettifor}, and the discussion of this problem in
Ref.~\cite{Hafner}.
The balance between 4$sp$ and 3$d$ electrons is the more influenced 
the larger the exchange
splitting fields are because the minority band 3$d$~level is shifted towards
the 4$s$, 4$p$ levels and the majority band 3$d$~level is shifted away. 
Only in first order of the splitting energy, 
we can expect that no change in the 
overall 4$s$, 4$p$
population happens, as is imposed by the choice of our 4$s$, 4$p$ chemical
potentials. Presently, the flow between 4$s$, 4$p$ and 3$d$ electrons cannot
be described with our model Hamiltonian as the electron-electron
interaction within the 4$s$, 4$p$ shell and between 4$sp$ and 
3$d$ is not included.

The exchange splittings not only determine the magnetic moment
but also influence strongly the shape of the single-particle bands in the
vicinity of $E_{\rm F}$ (not the overall bandwidth).
For the detailed comparison with photo-emission data we have thus either
chosen calculations with small $C$ values ($0.1\, {\rm eV}$), 
where the  minimum of $E_{\rm tot}(m)$
yields $m = 0.55\mu_{\rm B}$, or, for larger $C$~values, with a
fixed moment constraint, using the experimental spin-only moment of 
$\mu_{\hbox{\scriptsize spin-only}} = 0.55$.
The resulting quasi-particle bands do not differ much from each other. There
is however a tendency that values $C\approx 0.4\, {\rm eV}$ and larger
appear to agree somewhat better with the bulk of the photo-emission data.

Generally, the Gutzwiller results agree much better with experiment than
the SDFT results. For example, this is the case for, 
(i), the value for the quasi-particle density of states at the Fermi energy
($N_{\hbox{\scriptsize G-DFT}}^*(E_{\rm F})=2.6$ vs.~$3.0$ states/(eV atom)),
(ii), the positions of individual quasi-particle energies,
(iii), the values of the exchange splittings,
(iv), their $t_{2g}$-$e_g$~anisotropy, and, (v),
the $t_{2g}/e_g$~ratio of the $d$~part of the magnetic moment
($(t_{2g}/e_g)_{\hbox{\scriptsize G-DFT}}=83/17$ vs.~$81/19$). 
As a consequence of the 
small $d(e_g)$~exchange splitting, the $X_{2\downarrow}$
state lies {\sl below\/} $E_{\rm F}$ and, thus, the Fermi surface 
exhibits only {\sl one\/} hole
ellipsoid around $X$, in nice agreement with experiment. 

The large anisotropy of the exchange splittings is a result of our
ground-state energy optimization, which allows $\Delta_{t_{2g}}$ and 
$\Delta_{e_g}$ to be
independent variational parameters. We find
$\Delta_{t_{2g}} \approx 3\Delta_{e_g} \approx 800\, {\rm  meV}$. 
Note that these values enter
$|\Phi_0\rangle^{\rm bare}$ and are renormalized by factors 
$q_{d,\uparrow}\approx 0.7$, 
$q_{d,\downarrow}\approx 0.6$, 
when $|\Phi_0\rangle^{\rm opt}$ is reached.
This also implies that 
the width of the majority spin bands is about 10\% bigger than
that of the (higher lying) minority spin bands. It causes a further reduction
of the exchange splittings of states near $E_{\rm F}$, 
especially for those with strong $t_{2g}$ character. 
Note that this band dispersion effect causes larger exchange
splittings near the bottom of the $d$~bands, e.g., $0.45\, {\rm eV}$ 
splitting of $X_1$ and $0.74\, {\rm  eV}$ splitting of $X_3$. 
There, however, the quasi-particle linewidths have
increased to $1.25\, {\rm eV}$ and $1.4\, {\rm  eV}$, 
respectively~\cite{Eberhardt},
so that an exchange splitting near the bottom of the $d$ bands could, so far, 
not be observed experimentally.

The large anisotropy may originate from peculiarities
special to Ni with its almost completely filled $d$~bands and its fcc lattice
structure. Near the top of the $d$~bands, the $t_{2g}$~states dominate
because they exhibit the biggest hopping integrals to nearest neighbors,
$t^{(1)}_{dd\sigma}\approx 0.5\, {\rm eV}$. 
The $e_g$~states have $t^{(1)}_{dd\pi} \approx -0.3\, {\rm eV}$ to nearest
neighbors, and $t^{(2)}_{dd\sigma} \approx 0.1\, {\rm eV}$ to next-nearest
neighbors; the latter are small because of the large lattice distance
to second neighbors. The $e_g$~states also mix with the nearest-neighbor
$t_{2g}$~states with $t^{(1)}_{dd\pi}$-type coupling.
Therefore, the system can gain more band energy by avoiding occupation of
anti-bonding $t_{2g}$~states in the minority spin bands via large values of
$\Delta_{t_{2g}}$, at the expense of allowing occupation of 
less anti-bonding $e_g$ states via small $\Delta_{e_g}$~values. 
This scenario should not apply to materials with a bcc lattice structure 
which have almost equal nearest and next-nearest neighbor
separations. 
Since the bands in nickel are almost completely filled,
the suppression of charge fluctuations actually reduces
the number of atomic configurations where the Hund's-rule coupling
is active. It is also in this respect that
nickel does not quite reflect the generic situation of
other transition metals with less completely filled $d$~bands.

The results for nickel presented here must be seen as preliminary
inasmuch some important interaction terms were not yet included;
see Sect.~\ref{critics}. However, the present study already shows
that the Gutzwiller-DFT is a working approach.
It should allow us to resolve many of the open
issues in itinerant ferromagnetism in nickel and other transition metals.

\section{Conclusions and Outlook}

Which scenario for itinerant ferromagnetism in transition
metals is the correct one?

Band theory along the lines of Slater and Stoner could be worked out
in much detail whereas a correlated-electron description
of narrow-band systems was lacking until recently.
Our results for a two-band model and for nickel 
show that the van-Vleck scenario
is valid. Band theory alone does not account for the strong electronic
correlations present in the material which lead to the observed
renormalization of the effective mass, exchange splittings, bandwidths,
and Fermi surface topology. Moreover,
charge fluctuations are indeed small, and large local moments are present
both in the paramagnetic and the ferromagnetic phases.

Roughly we may say that the electrons' motion 
through the crystal leads to a ferromagnetic
coupling of pre-formed moments which eventually
order at low enough temperatures.
In this way, strong itinerant ferromagnets resemble ferromagnetic
insulators as far as their low-energy properties are
concerned: spinwaves exist which destroy the magnetic
long-range order at the Curie temperature. 

Our present scheme allows us a detailed comparison with data
from refined photo-emission experiments on nickel which are currently
carried out~\cite{Claessen}. It should be clear that our approach
is applicable not only to nickel but to all other itinerant
electron systems.

Despite all recent progress much work remains to be done. The
present implementation of the Gutzwiller-DFT needs to be improved
by the inclusion of more orbits, their mutual Coulomb interaction terms,
and the spin-orbit coupling. Ultimately, some of the principle limitations
of our variational approach will have to be overcome by a fully dynamic 
theory. Most probably, such a theory will require enormous numerical
resources such that a fully developed Gutzwiller-DFT will always remain a 
valuable tool to study ground-state properties of correlated electron
systems.

\section*{Acknowledgments}

We gratefully acknowledge helpful discussions with 
all participants of the Heraeus seminar {\sl Ground-State and
Finite-Temperature Bandferromagnetism\/}.
This project is supported in part by
the Deutsche Forschungsgemeinschaft under WE~1412/8-1.

%INDEX%%%%%%%%%%%%%%%%%%%%%%%%%%%%%%%%%%%%%%%%%%%%%%%%%%%%%%%%%%%%%%%
% Please check with the editor of your book whether he plans to
% include a "mutual" subject index - if so, please code your entries
% in the standard syntax. For your own purposes you may print your
% "personal" index by using the following commands:
%
%\clearpage
%\addcontentsline{toc}{section}{Index}
%\flushbottom
%\printindex
%%%%%%%%%%%%%%%%%%%%%%%%%%%%%%%%%%%%%%%%%%%%%%%%%%%%%%%%%%%%%%%%%%%%%


\begin{thebibliography}{99.}
\bibitem{hint} This holds for our symmetry-restricted basis.
\end{thebibliography}
\end{document}